\documentclass[%
 aip,
 amsmath,amssymb,
 reprint,%
]{revtex4-1}

\usepackage{graphicx}
\usepackage{dcolumn}
\usepackage{bm}

\usepackage[utf8]{inputenc}
\usepackage[T1]{fontenc}
\usepackage{mathptmx}
\usepackage{etoolbox}
\usepackage{subfigure}

\makeatletter
\def\@email#1#2{%
 \endgroup
 \patchcmd{\titleblock@produce}
  {\frontmatter@RRAPformat}
  {\frontmatter@RRAPformat{\produce@RRAP{*#1\href{mailto:#2}{#2}}}\frontmatter@RRAPformat}
  {}{}
}%
\makeatother
\begin{document}

\preprint{AIP/123-QED}

\title[]{Ultra-high vacuum pressure measurement using cold atoms}

\author{S. Supakar}
\affiliation{  Laser Physics Applications Division, Raja Ramanna Centre for Advanced Technology, Indore-452013, India }

\author{Vivek Singh*}
\affiliation{  Laser Physics Applications Division, Raja Ramanna Centre for Advanced Technology, Indore-452013, India }
\affiliation{Homi Bhabha National Institute, Anushaktinagar, Mumbai-400094, India }
\email{viveksingh@rrcat.gov.in}

\author{V. B. Tiwari}
\affiliation{  Laser Physics Applications Division, Raja Ramanna Centre for Advanced Technology, Indore-452013, India }
\affiliation{Homi Bhabha National Institute, Anushaktinagar, Mumbai-400094, India }

\author{S. R. Mishra}

\affiliation{  Laser Physics Applications Division, Raja Ramanna Centre for Advanced Technology, Indore-452013, India }

\affiliation{Homi Bhabha National Institute, Anushaktinagar, Mumbai-400094, India }

\date{\today}

\begin{abstract}
 In this work, we have measured the background pressure in an ultra-high vacuum (UHV) chamber by measuring the collisional loss rates in a Rb atom magneto-optical trap (MOT) on an atom chip. The loss rate due to non-Rb gases in the background has been estimated by measuring the MOT loss rate in low Rb pressure regime. These results can be useful for development of cold-atoms based UHV pressure standards.
\end{abstract}

\maketitle
Magneto-optical trap (MOT) is a robust device to generate the samples of ultracold atoms for various research and device applications of these cold atoms. Nowadays, cold atoms are considered as an important quantum systems for their applications in several upcoming quantum technologies such as high precision atomic clocks \cite{Wang, Guena}, inertial sensors \cite{Geiger, Nelson, chu, wu, lee, bidel,alzar}, electro-magnetic field sensors \cite{carter, wild, beh}, quantum computers \cite{briegel, saffman}, etc. Recently, the use of cold atoms for developing quantum vacuum pressure standard has been proposed and demonstrated \cite{shen1, shen2}. The cold atom based pressure sensors are absolute and universal, as they are based on atomic collision process and no repeated calibration is required over the time. This is advantage of a cold atom based pressure standard over the conventional pressure sensing instruments such as ionization gauges which require repeated calibrations due to aging of filaments and electrodes. In addition, the cold atoms based pressure standards can work over the large dynamic range of vacuum, from UHV to extreme-high vacuum (XHV) regime.

The loss rate of atoms in atom traps are sensitive to background pressure in the trap chamber \cite{arpo, eckel, eckelrsi, moore, yuan, wu1, will, vivek3}. Therefore atom traps can be utilized to sense or measure the UHV pressure in the chamber. Both, MOT and magnetic trap, are being used for pressure sensing applications with their relative advantages over each other. MOT is easier to form but it can sense pressure in UHV regime only, whereas the magnetic traps can be used to sense the pressure down to XHV regime. Earlier, Yuan et al \cite{ yuan} have estimated the Rb-pressure in the chamber from the MOT loading time in low cooling beam intensity regime by ignoring the intra-trap collisional loss rate. Willems et al \cite{will} have estimated the background pressure (non-Cs gasses) in the chamber by measuring the life-time of MOT and magneto-static trap. Arpornthip et al \cite{arpo} measured the MOT loading time as function of background pressure (non-Rb contents) as well as the function of MOT loading rate (dependent on Rb-pressure), to estimate the background pressure and partial pressure of Rb in the chamber. Moore et al \cite{moore} has measured the non-Rb background pressure in the chamber from the Rb-MOT loading data by increasing the non-Rb gas pressure in the chamber - applying an approach of chamber pressure rise demonstrated earlier by Arpornthip et al. In this method, sputter ion pump (SIP) was turned off to change the non-Rb gas pressure and MOT loading was studied at different non-Rb background pressure. Though this method is more time consuming, but it is suitable to detect vacuum leak in the chamber. In an another approach \cite{vivek3}, the partial pressure due to Rb and non-Rb gases have been estimated by measuring the saturated number and loading time in a Rb-MOT.  \\

In the work reported here, we have estimated the background pressure due to non-Rb gases in the chamber by measuring the Rb-MOT loading time in low Rb pressure and low cooling beam intensity regimes. We first measured the MOT loss rate ($\Gamma $) as function of cooling beam intensity. The MOT loading time at low cooling beam intensity was used to estimate the total (Rb and non-Rb) background collisional loss rate by neglecting the intra-trap collisional loss rate. Then, we measured the loading time of this low cooling beam intensity MOT as function of Rb-dispenser current. In these measurements, the MOT loading time at low dispenser current was used to estimate the MOT loss rate due to non-Rb background gas contents, which provided the estimate of non-Rb background pressure in the chamber. Therefore, as compared to earlier methods \cite{ yuan, arpo, moore},  we show that non-Rb partial pressure in the UHV chamber can be estimated by measuring the MOT loading time in low cooling beam intensity and low Rb pressure regimes. Our method is comparatively less time consuming and does not require switching-off the pumping of vacuum chamber which prevents exposure of the chamber to undesirable gas contamination. The straight forward method presented here has the potential for developing a UHV pressure sensor device.\\

The MOT loading process can be described by a rate equation as \cite{arpo},
\begin{equation}
\frac{dN(t)}{dt}= R - \gamma_{b} N(t) - \beta \bar{n}(t)N(t)    
\end{equation}
where $N (t)$ is the number of atoms in the MOT cloud at any time $t$, $R$ is the loading rate of MOT due to Rb vapour in the background, $\gamma_{b}$ is the loss rate in MOT due to collisions of trapped atoms in MOT with the atoms/molecules present in the background, $\beta$ is loss rate due to inelastic two body intra-trap collisions, $\bar{n}(t) = \int_{}^{} n(\textbf{r},t)^2 \,dV / N(t) $ is average number density  and $n(\textbf{r},t)$ the number density of the trapped atoms in the MOT cloud. \\

The solution of equation (1) depends on the regime of parameter in which MOT is operated. For small number of atoms in MOT ($N<10^5$), known as constant volume regime, $\bar{n}(t)\approx {N(t)}/{V}$. For large N ($N>10^5$), known as constant density regime,  $\bar{n}$ is constant. In our experiments, the  MOT was operated in the constant density regime (i.e. $N>10^6$), therefore the solution of the equation (1) can be written as,
\begin{equation}
N(t) = N_{s} \left[1-exp(-t/\tau_{L})\right],
\end{equation}
where $\tau_{L}$= 1/$\Gamma$ with $\Gamma = \gamma_{b} + \beta \bar{n}$. Here $N_{s} = R\tau_{L}$ is the final number in the MOT (i.e. number of atoms in the MOT in equilibrium). The parameter $\tau_{L}$ is known as MOT loading time. The equation (2) describes the variation in number of atoms in a MOT with time and its plot is referred as MOT loading curve. From the experimentally measured MOT loading curve, both parameters, $\tau_{L}$ and $N_{s}$, can be determined. These parameters are dependent on the background pressure in the chamber due to Rb and non-Rb gas contents, and on the MOT parameters such as cooling beam intensity etc. 

It is known that the loss rate of atoms from the MOT cloud due to collisions from atoms/molecules of any gas species in the background is related to its partial pressure in the chamber. The loss rate $\gamma_{i}$ due to collisions with atoms/molecules of $i^{th}$ gas species in the background is related to its partial pressure $P_{i}$ in the background as \cite{arpo},
\begin{equation}
\gamma_{i} = 6.8\frac{P_{i}}{(k_{B}T)^{2/3}}\left(\frac{C_{i}}{m_{i}}\right)^{1/3} (Dm_{0})^{-1/6} = \frac{P_{i}}{k_{i}}, 
\end{equation}
where $m_{0}$ is mass of atom in the trap, $m_{i}$ is mass of the incident atom/molecule of $i^{th}$ gas species, $k_{B}$ is Boltzmann constant, T is the temperature and D is the trap depth of the MOT and $C_{i}$ is Van der Walls coefficient for  $i^{th}$ gas species in the background. \\

In the typical vapour loaded MOT chamber, the background collisional loss rate has two components and can be written as, 
\begin{equation}
\gamma_{b} =\gamma_{Rb} + \gamma_{non-Rb},
\end{equation}
where $ \gamma_{Rb}$($= (k_{Rb})^{-1} P_{Rb}$)  represents the loss rates due to collisions with untrapped Rb vapour atoms and $\gamma_{non-Rb}$ represents the loss rate due to other atoms/molecules in the background.

\begin{figure}[h!]
\centering
\subfigure[]{
\label{gamma vs int}
\resizebox*{8.7 cm}{!}{\includegraphics{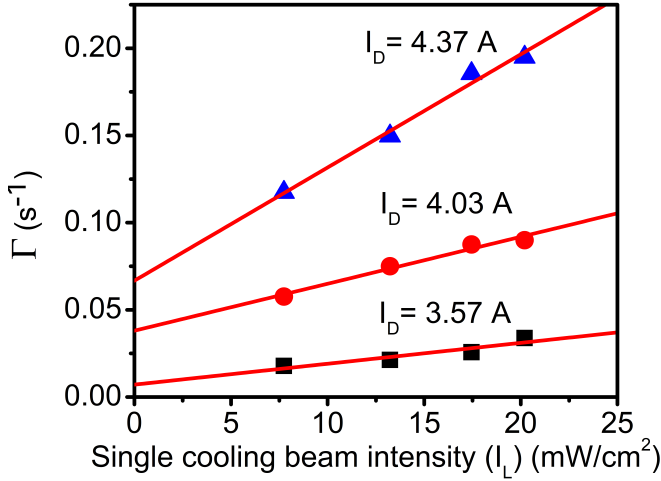}}\hspace{0 cm}}
\subfigure[]{
\label{gammavsdispenser}
\resizebox*{9.0 cm}{!}{\includegraphics{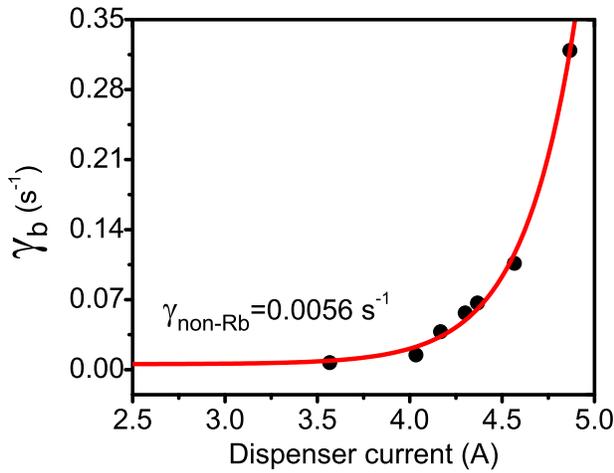}}}
\caption{(a) The variation in the loss rate ($\Gamma$) with cooling laser beam intensity for different Rb dispenser current ($I_{D}$). (b) Variation in background collisional loss rate ($\gamma_{b}$) with Rb dispenser current.} \label{sample-figure}
\end{figure}

We have experimentally measured $\Gamma$ for different values of laser beam intensity in the MOT. In the low intensity regime of cooling laser beam, the value of  $\Gamma$ can be approximated to $\Gamma \, \approx \, \gamma_{b}$, where the intratrap collisional loss rate ($ \beta \bar{n} $) is negligible as compared to background collisional loss rate. For value of $\bar{n} \sim 10^{8} \,$ cm$^{-3}$ (for our MOT at 7.7 mW/cm$^{2}$) and $ \beta \sim 2 \times 10^{-12} \,$ cm$^{-3}$ s$^{-1}$ (as reported earlier \cite{Gensemer} for detuning and intensity used in our MOT), the value of $ \beta \bar{n} $ is $\sim 10^{-4} \,$ s$^{-1}$. This is much smaller than the lowest value of $\gamma_{b} $ ($\sim 0.0071 \,$ s$^{-1}$) observed at that intensity (Figure \ref{gamma vs int}).

\begin{figure}[ht]
\centering\includegraphics[width=8.5cm]{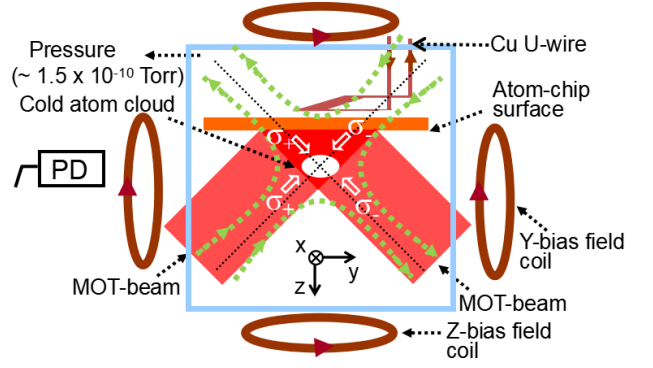}
\caption{A schematic diagram of the experimental setup. Two MOT-beams in the reflection geometry in the y-z plane are shown, whereas the other two MOT-beams along $\pm$ x-direction are not shown in the diagram. PD represents photodiode (PD) for the detection of fluorescence.}
\label{exptsetup}
\end{figure}

The experiments have been performed with loading of mirror-MOT (U-MOT) on an atom-chip, with schematic as shown in Figure \ref{exptsetup}. The details of experimental setup of atom-chip mirror-MOT have been described earlier \cite{ vivek2, vivek3}. Different vacuum pumps used in the setup include a 77 l/s turbo molecular pump (TMP), a 300 l/s sputter ion pump (SIP) and a titanium sublimation pump (TSP). The ultimate base pressure achieved in the chamber without Rb vapour was $1.5 \times 10^{-10}$ Torr as read by SIP controller. The pressure values read by SIP controller were nearly equal to those read by an extractor gauge attached to the chamber. A Rb dispenser assembly having three Rb dispensers (Rb/NF/3.4/12FT) connected in parallel configuration was prepared by welding dispensers on a two-pin feedthrough. This assembly was placed in the vacuum chamber through a viewport hole such that a Rb dispensers are at a distance of $\sim$ 17 cm from the centre of the octagonal chamber. The Rubidium vapour is produced inside the chamber after flowing a current through this dispenser assembly. The current in each dispenser ($I_{D}$) is nearly one-third of the current supplied to dispenser assembly.

\begin{figure}[h]
\centering\includegraphics[width=8.5cm]{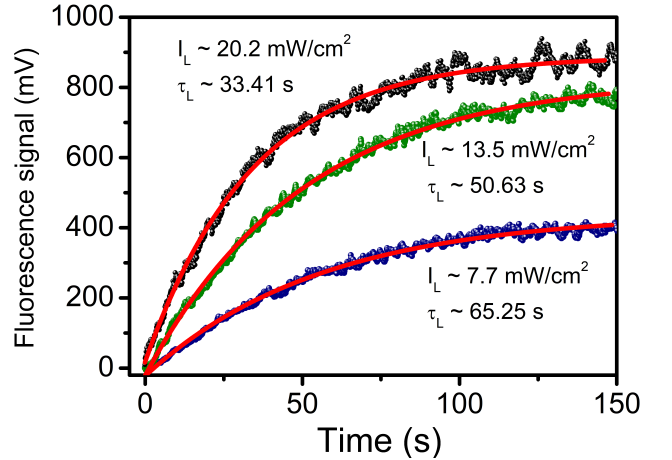}
\caption{The loading curves for U-MOT on atom-chip for different values of cooling laser beam intensity at a fixed background pressure (at dispenser current of $I_{D}$ = 3.57 A). The experimentally observed MOT loading data along with best-fit (continuous curve) are shown for different values of intensity.}
\label{loadingcurve}
\end{figure}

A quadrupole like magnetic field required for MOT was generated from a current carrying (60 A) copper U-wire (Figure \ref{exptsetup}) placed behind the atom-chip in presence of a homogeneous bias fields ($B_{y} \, \sim$  11 G and $B_{z} \, \sim$  3 G) . The output from frequency stabilized diode lasers served as cooling and repumping laser beams. Each MOT beam was a combination of a cooling and a re-pumping beams with suitable ratio of power in the beams. Two MOT beams were reflected at 45$^{\circ}$ from chip surface which formed four MOT beams in the overlapping region. Two counter propagating MOT beams in orthogonal direction made a set of required six MOT beams for operation of the MOT. This MOT configuration is called the mirror-MOT configuration. 

Figure \ref{loadingcurve} shows the loading curve of U-MOT for different values of cooling laser beam intensity at a fixed Rb dispenser current of $I_{D}$ = 3.57 A. The continuous curves show the best-fit of the experimental loading curve to the equation (2). From the fit, we obtain the value of loading time $\tau_{L}$ and $\Gamma$. A reduction in loading time from 65.25 s to 33.41 s was observed with increase in intensity of the cooling beam from 7.7 mW/cm$^{2}$ to 20.2 mW/cm$^{2}$, as shown in figure \ref{loadingcurve}. These intensity dependent measurements of $\Gamma$ were carried out for different values of dispenser current ($I_{D}$) and results are shown in Figure \ref{gamma vs int}. As discussed before, the value of $\Gamma ( = \gamma_{b} + \beta \bar{n})$ in low cooling beam intensity regime can be approximated as $\Gamma \, \sim \,\gamma_{b}$. Alternatively, as followed by Yuan et al \cite{yuan}, the intercept on y-axis in $\Gamma$ vs cooling beam intensity plot. Figure \ref{gamma vs int} can be used to estimate the value of $\gamma_{b}$. Figure \ref{gammavsdispenser} shows the $\gamma_{b}$ values estimated this way for different values of dispenser current.   

As shown in figure \ref{gammavsdispenser}, the value of $\gamma_{b}$ increases rapidly with dispenser current for current beyond the value of 4.0 A. However, at lower dispenser current (lower than 4.0 A) values, the variation in $\gamma_{b}$ with current is negligibly small. This shows that contribution to loss rate from the Rb atoms in the background is negligible. Therefore, the value of $\gamma_{b}$ can be approximated to $\gamma_{non-Rb}$ in this regime. By estimating $\gamma_{non-Rb}$ this way, $\gamma_{Rb}$ can be estimated at any current value by measuring $\gamma_{b}$, as $\gamma_{non-Rb}$ is independent of dispenser current. Thus, we can estimate both $\gamma_{non-Rb}$ and $\gamma_{Rb}$ in our method, without switching off the vacuum pumps as compared to earlier works\cite{arpo, moore}.

In the very low pressure (UHV) regime, there are only few gas species (H$_{2}$, He, Ar etc) which contribute to the total base pressure in the chamber ($P = \sum_{i} P_{i} = \sum_{i} k_{i} \gamma_{i}$). If we consider hydrogen (H$_{2}$) as a dominant species in the UHV pressure range as in our chamber \cite{red} , the measured $\gamma_{non-Rb}$ can be approximated to $\gamma_{H_{2}}$.  This gives the non-rubidium partial pressure in the chamber as $P_{non-Rb} = k_{H_{2}} \, \gamma_{H_{2}} = \, k_{H_{2}} \, \gamma_{non-Rb}$ =  $1.1 \times 10^{-10} \,$ Torr, with $k_{H_{2}} = 2.04 \times 10^{-8} $ Torr s and  $\gamma_{non-Rb}$ = 0.0056 s$^{-1}$. This estimated pressure of non-Rb background gases agrees well with that measured by the SIP controller ($ 1.5 \times 10^{-10}$ Torr).

\begin{figure}[h]
\centering\includegraphics[width=9.5cm]{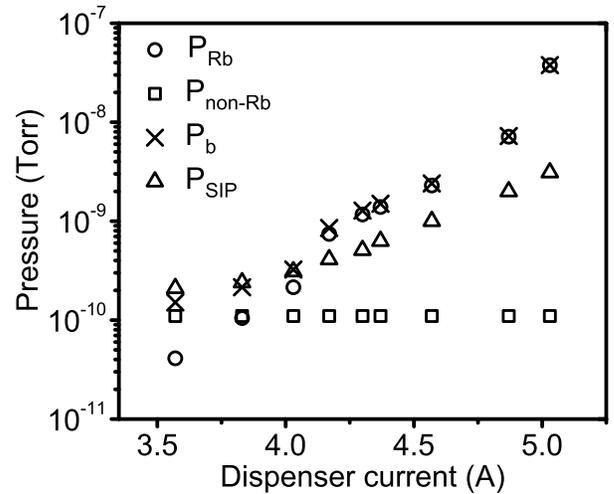}
\caption{Variation in partial pressures due to Rb vapour ($P_{Rb}$), non-rubidium gases ($P_{non-Rb}$). The total background pressure ($P_{b}$ \, = \,$P_{Rb}$ + $P_{non-Rb}$) is compared with pressure measured by SIP controller ($P_{SIP}$).}
\label{pressures vs dispenser}
\end{figure}

After knowing the value of $\gamma_{Rb}$ at any dispenser current, the value of Rb partial pressure can be estimated by the relation $P_{Rb} = k_{Rb} \gamma_{Rb}$, with $k_{Rb} = 2.27 \times 10^{-8}\,$ Torr s \cite{yuan, arpo}. The variation of the estimated Rb pressure in the chamber with dispenser current is shown in figure \ref{pressures vs dispenser}. In this figure, the total background pressure ($P_{b} \, = \, P_{Rb} \, + \, P_{non-Rb}$) estimated by our method is also shown and compared with the pressure read by SIP controller. We note that at low dispenser current (less than 4.0 A), there is a good agreement between the estimated total background pressure ($P_{b}$) and the pressure measured by SIP controller ($P_{SIP}$). However, at higher values of dispenser current, the pressure estimated by present method is more than that read by SIP controller. This difference can be attributed to the adsorption of Rb atoms at the chamber walls and pipe connecting the SIP to the chamber. Similar observations have been reported earlier also \cite{arpo}.


In conclusion, we have estimated the Rb and non-Rb partial pressure values in an UHV chamber from the loading data of a Rb-MOT on an atom-chip. The estimated pressure values agree with the pressure measured by the SIP controller.
\\ 

We are thankful for the help extended by Amit Chaudhary and Dayanand Mewara for this work. \\

\noindent The authors declare no conflict of interest.
\bibliography{reference}

\end{document}